\documentclass[alpha-refs]{wiley-article}

\usepackage{color}
\usepackage{ulem}

\usepackage{siunitx}
\usepackage{comment}

\papertype{Original Article}
\paperfield{Journal Section}

\title{Non-conservation and conservation for different formulations of moist potential vorticity}

\author[1]{Parvathi Kooloth}
\author[2]{Leslie M. Smith}
\author[2,3]{Samuel N. Stechmann}

\affil[1]{Atmospheric, Climate, \& Earth Sciences Division, Pacific Northwest National Laboratory, Richland, WA, 99354, USA}
\affil[2]{Department of Mathematics, University of Wisconsin--Madison, Madison, WI, 53706, USA}
\affil[3]{Department of Atmospheric and Oceanic Sciences, University of Wisconsin--Madison, Madison, WI, 53706, USA}

\corraddress{Parvathi Kooloth, Atmospheric Sciences and Global Change Division, Pacific Northwest National Laboratory, Richland, WA, 99354, USA}
\corremail{parvathi.kooloth@pnnl.gov}

\fundinginfo{US NSF DMS grant number 1907667 \\[24pt] 
Submitted on November 15, 2023}

\runningauthor{Kooloth et al.}

\begin{document}
\maketitle
\begin{abstract}
Potential vorticity (PV) is one of the most important quantities in atmospheric science. The PV of each fluid parcel is known to be conserved, in the case of a dry atmosphere. However, a parcel's PV is not conserved if clouds or phase changes of water occur. Recently, PV conservation laws were derived for a cloudy atmosphere, where each parcel's PV is not conserved but parcel-integrated PV is conserved, for integrals over certain volumes that move with the flow. Hence a variety of different statements are now possible for moist PV conservation and non-conservation, and in comparison to the case of a dry atmosphere, the situation for moist PV is more complex. Here, in light of this complexity, several different definitions of moist PV are compared for a cloudy atmosphere. Numerical simulations are shown for a rising thermal, both before and after the formation of a cloud. These simulations include the first computational illustration of the parcel-integrated, moist PV conservation laws. The comparisons, both theoretical and numerical, serve to clarify and highlight the different statements of conservation and non-conservation that arise for different definitions of moist PV. 
\end{abstract}
\newpage
\section{Introduction}
\label{sec:intro}

Potential vorticity (PV) is one of the central conserved quantities in geophysical fluid dynamics \citep{muller1995ertel,salmon1998lectures}, with its roots traced back to about a century ago in the works of \citet{rossby1939relation} and \citet{ertel1942neuer}. The PV conservation law also has a deep connection to the classic Kelvin and Bjerknes' circulation theorems \citep{bjerknes1898hydrodynamischen,thomson1867,thorpe2003bjerknes}.
Pointwise conservation enables the use of PV as a tracer of fluid parcels. PV also possesses an inversion principle that allows one to  recover the slowly varying component of the wind and temperature fields from the PV distribution with appropriate boundary conditions \citep{hoskins1985use,martin2013mid}. Owing to these properties, PV has been used extensively to study the dynamics of synoptic and mesoscale weather systems \citep{hoskins1985use,thorpe1985diagnosis, davis1991potential, lackmann2002cold} and also ocean circulations \citep{holland1984dynamics, rhines1986vorticity, pollard1990large, marshall1992fluid, taylor2010buoyancy, ruan2021evolution}.

%

Moist versions of PV have also been proposed and investigated extensively \citep{bennetts1979conditional,emanuel1979inertial,schubert2001potential,marquet14definition,wetzel2019balanced,wetzel2020potential}.
Moist PV has been useful for many purposes, including the diagnosis of the effects of latent heating
\citep{bennetts1979conditional,emanuel1979inertial,davis1991potential,cao1995generation,lackmann2002cold,gao2004generation,brennan2005influence,korty2007climatology,emanuel2008back,martin2013mid,lackmann2011midlatitude,madonna2014warm,bueler2017potential,hittmeir2021dynamic,abbott2024impact}.


Note, though, that moist PV is not conserved for each fluid parcel, 
and inversion of moist PV is problematic \citep[e.g., see the sequence of three studies of][]{cao1995generation,schubert2001potential,wetzel2020potential}. 
The traditional conservation and inversion properties of dry PV are  for idealized single-component flows, not for the more realistic cases of binary or multi-component fluids such as a moist atmosphere with clouds and phase changes, and salty oceans.

For inversion, the moist case is different than the dry case in several ways. For instance, the balanced portion is comprised of not one but two components (PV and M, where M represents a slow moist component), and correspondingly it is not PV inversion but PV-and-M inversion that recovers the balanced portion of the system \citep{smith2017precipitating,wetzel2019balanced,wetzel2020potential,remond2023nonlinear}.
Also, among many different moist PV quantities that have been used, only certain moist PV quantities are slowly evolving in the presence of phase changes of water and cloud latent heating \citep{wetzel2020potential,zhang2021effects,zhang2021fast,zhang2022convergence}.

For conservation, recently, we generalized the PV conservation laws to cases with phase changes of water, for both a  compressible flow \citep{kooloth2022conservation} and a Boussinesq flow \citep{kooloth2023hamilton}. We showed that moist PV is not pointwise conserved as in a dry atmosphere; instead it is conserved over certain `material' volumes that move with flow. Such conservation laws hold for many, but not all, moist PV quantities.

The purpose of this letter is to 
present a detailed comparison of different statements of PV conservation and  
non-conservation for various definitions of moist PV.
One part of this comparison is the first numerical illustration of the parcel-integrated PV conservation law. From these comparisons, we hope to bring some clarity to the complex landscape of cases including dry versus moist PV, conservation and non-conservation, and parcel-wise versus parcel-integrated conservation.

In what follows, the equations of PV conservation and non-conservation are described for a compressible atmosphere (section~\ref{sec:background}) and under the Boussinesq approximation (section~\ref{sec:num-sim}). The setup of the numerical simulation and the simulation results are also presented in section \ref{sec:num-sim}. Finally, section~\ref{sec:conclusions} includes a concluding discussion and summary of the laws of conservation and non-conservation.







\section{Comparison of PV conservation laws}
\label{sec:background}

In this section, we compare a variety of statements of 
conservation and non-conservation of moist potential vorticity,
including recently discovered conservation laws that apply
for binary or multi-component fluids such as an ocean with salinity
or an atmosphere with water vapor, and even in the presence of phase changes
and clouds
\citep{kooloth2022conservation,kooloth2023hamilton}.

The dry case without water vapor is considered in section~\ref{sec:dry-pv}, 
where PV is conserved for each fluid parcel. 
Then the moist case with clouds and phase changes 
is considered, 
where each fluid parcel's PV is not conserved (section~\ref{sec:moist-pv-point}),
but where a local-volume-integrated PV is conserved
(section~\ref{sec:moist-pv-int}).

The setting in this section is a compressible atmosphere.
See section~\ref{sec:num-sim} below for an alternative setting under the Boussinesq approximation.





For the evolution equations and assumptions,
for velocity
$\vec{u}=(u,v,w)$ and density $\rho$,
the form of the equations is the same for
both the dry and moist cases:
\begin{equation}
    \frac{D\vec{u}}{Dt} = - \frac{1}{\rho} \nabla p + \nabla \phi,
    \quad
    \frac{D\rho}{Dt} = -\rho\nabla\cdot\vec{u},
    \label{eqn:momentum}
\end{equation}
where $\nabla=(\partial_x,\partial_y,\partial_z)$ is the gradient operator,
$D/Dt = \partial/ \partial t + \vec{u} \cdot \nabla $ is the material derivative, $p$ is the pressure and $\phi$ is the force potential, which could include, for instance, the gravitational potential.
In (\ref{eqn:momentum}) a case without dissipation (friction, viscosity, etc.) is assumed.
Rotation could be added with some modifications \citep{kooloth2022conservation}
but is left out for simplicity here.
Also, in (\ref{eqn:momentum}), the density $\rho$ and pressure $p$
should be interpreted as 
total density and total pressure, for the dry case and also for
the moist case, so that the form of (\ref{eqn:momentum}) is the same 
in both cases.
To complete the specification of the dynamical evolution,
equations are also needed for thermodynamic quantities.

For a dry atmosphere, the thermodynamic evolution equation can be described in terms of potential temperature $\theta$ as
\begin{equation}
    \frac{D\theta}{Dt} = 0,
    \label{eqn:theta-evol}
\end{equation}
and an equation of state, $\theta=\theta(p,\rho)$.
The evolution is assumed to be (dry) adiabatic.

For a moist (and possibly cloudy) atmosphere, the thermodynamic evolution equations can be described in terms of 
equivalent potential temperature $\theta_e$ and
total water specific humidity $q_t$ as
\begin{equation}
    \frac{D\theta_e}{Dt} = 0,
    \quad
    \frac{Dq_t}{Dt} = 0,
    \label{eqn:thetae-qt-evol}
\end{equation}
along with an equation of state, $\theta_e=\theta_e(p,\rho,q_t)$.
The evolution is assumed to be (moist) adiabatic,
with reversible phase changes between water vapor and liquid water,
as in warm, liquid clouds.
No rain, ice, nor precipitation are considered.
The total water $q_t$ can be decomposed as
$q_t=q_v+q_l$, where $q_v$ and $q_l$ could be recovered from
$q_t$ by comparison against the saturation specific humidity,
$q_{vs}=q_{vs}(T,p)$,
where $T$ is temperature.
We refer to states with $q_t<q_{vs}$ and $q_t\ge q_{vs}$
as the unsaturated and saturated phases, respectively.

In what follows, we consider
various definitions of moist PV that have been proposed in the past.
All definitions involve the vorticity, $\Vec{\omega}=\nabla\times\vec{u}$.
In analogy with dry PV, $(\vec{\omega}\cdot\nabla\theta )/\rho$,
the various moist PV definitions are distinguished by which 
thermodynamic quantity is used in 
$\vec{\omega}\cdot\nabla\psi$, where common choices of $\psi$ include
potential temperature $\theta$,
virtual potential temperature $\theta_v$,
equivalent potential temperature $\theta_e$,
or
liquid water potential temperature $\theta_l$.

\subsection{Dry PV is conserved for each fluid parcel}
\label{sec:dry-pv}

Dry PV, $(\vec{\omega}\cdot\nabla\theta )/\rho$,
is a material invariant---i.e., conserved
for each fluid parcel. To see this, the starting point is its evolution
equation:
\begin{align}
    \rho \frac{D}{Dt} \left( \frac{\vec{\omega} \cdot \nabla \theta} {\rho} \right) &=  -\nabla \theta \cdot \nabla \times \left( \frac{1}{\rho} \nabla p\right) \nonumber \\ 
    &=  -\nabla \theta \cdot \left( \nabla \frac{1}{\rho} \times \nabla p\right),
    \label{eqn:pv-evol-dry}
\end{align}
which can be derived by using 
(\ref{eqn:momentum}) and (\ref{eqn:theta-evol})
\citep[see, e.g.,][]{kooloth2022conservation}.

To see that the ``solenoidal'' term on the right-hand side of
(\ref{eqn:pv-evol-dry}) is zero for a dry atmosphere,
recall a fundamental property of thermodynamics for a dry atmosphere: 
the potential temperature (or any other thermodynamic property) can be expressed as a function of pressure and density only, so that $\theta=\theta(p,\rho)$.
Consequently, we have
\begin{equation}
    \nabla \theta 
    = \frac{\partial \theta}{\partial p}\nabla p 
    +\frac{\partial \theta}{\partial \rho}\nabla \rho,
\end{equation}
and it follows that the right-hand side of (\ref{eqn:pv-evol-dry}) is zero,
so that
\begin{equation}
    \frac{D}{Dt} \left( \frac{\vec{\omega} \cdot \nabla \theta} {\rho} \right) = 0.
\end{equation}
Hence, dry PV is conserved for each fluid parcel.



\subsection{Moist PV is not conserved for each parcel, due to clouds}
\label{sec:moist-pv-point}

We now describe several common choices of moist PV definitions
(based on $\theta$, $\theta_v$, $\theta_e$, and $\theta_l$),
and show how each one is not a material invariant, in the presence 
of clouds and phase changes.

As a way to encapsulate any definition of moist PV,
consider a moist PV defined as $(\vec{\omega}\cdot\nabla\psi )/\rho$,
for a generic thermodynamic quantity $\psi$.
Assume that the evolution of $\psi$ is given by
$D\psi/Dt=\dot{\psi}$, where $\dot{\psi}$ represents
all sources/sinks of $\psi$.
Then the evolution of the generic moist PV is given by
\begin{align}
     \frac{D}{Dt} \left( \frac{\vec{\omega} \cdot \nabla \psi} {\rho} \right) =  \frac{1}{\rho^3} \nabla \psi \cdot \left( \nabla {\rho} \times \nabla p\right) + \frac{1}{\rho}\vec{\omega} \cdot \nabla \dot{\psi} 
    \label{eqn:pv-evol-psi}
\end{align}
which follows from (\ref{eqn:momentum}) 
\citep[see, e.g.,][]{schubert2001potential,kooloth2022conservation}.
Two potential sources of non-conservation appear on the right-hand
side in (\ref{eqn:pv-evol-psi}): the ``solenoidal'' term
involving a cross product and the source term involving $\dot{\psi}$.


First choose $\psi=\theta$ and
consider $PV_{\theta}$ based on potential temperature.
From (\ref{eqn:pv-evol-psi}) its evolution equation is
\begin{align}
     \frac{D}{Dt} \left( \frac{\vec{\omega} \cdot \nabla \theta} {\rho} \right) =  \frac{1}{\rho^3} \nabla \theta \cdot \left( \nabla {\rho} \times \nabla p\right) + \frac{1}{\rho}\vec{\omega} \cdot \nabla \dot{\theta}.
    \label{eqn:pv-evol-theta}
\end{align}
In an atmosphere with clouds and phase changes, 
(\ref{eqn:pv-evol-theta}) cannot be further simplified.
The $\dot{\theta}$ term arises from cloud latent heating
and does not vanish, and the solenoidal term on the right hand side of (\ref{eqn:pv-evol-theta}) remains nonzero in (\ref{eqn:pv-evol-theta}), since, for a moist atmosphere with phase changes,
the potential temperature is no longer completely determined by pressure and density. Consequently, 
$PV_\theta$ is not a material invariant if clouds and
phase changes are present.


Next consider $PV_{\theta_v}$ based on the virtual potential temperature $\theta_v$. Its evolution equation is
\begin{equation}
     \frac{D}{Dt} \left( \frac{\vec{\omega} \cdot \nabla \theta_v} {\rho} \right) = \frac{1}{\rho} \vec{\omega} \cdot \nabla \dot{\theta}_v,
     \label{eqn:PV_v-evolution}
\end{equation}
which follows from choosing $\psi=\theta_v$ in (\ref{eqn:pv-evol-psi}).
The solenoidal term has vanished and does not appear in (\ref{eqn:PV_v-evolution})\footnote{
The solenoidal term from (\ref{eqn:pv-evol-psi}) has vanished in (\ref{eqn:PV_v-evolution}) because $\theta_v$ is a function of $p$ and $\rho$ alone. Recall that virtual potential temperature can be expressed as $\theta_v = T_v (p_0/p)^{R_d/c_{p_d}}$, in terms of virtual temperature $T_v = p/(\rho R_d)$, and total pressure $p$. The other terms are constants: reference pressure $p_0$, gas constant $R_d$ for dry air and specific heat $c_{p_d}$ at constant pressure for dry air. },
which is one of the desirable properties of $PV_{\theta_v}$
\citep{schubert2001potential}.
The right hand side of ($\ref{eqn:PV_v-evolution}$) still has a source term resulting from non-conservation of $\theta_v$, such as cloud latent heating, so that $PV_{\theta_v}$ is not a material invariant in general. As a special case, though, while a parcel remains in the unsaturated phase, we have $\dot{\theta}_v=0$ due to the absence of latent heating, and consequently $PV_{\theta_v}$ remains materially conserved in the unsaturated phase.

Another commonly used definition of moist potential vorticity is $PV_{\theta_e}$ defined in terms of the equivalent potential temperature $\theta_e$. Note that $\theta_e$ is a materially conserved quantity, i.e., $D\theta_e/Dt=0$. Hence, from (\ref{eqn:pv-evol-psi}), the evolution of $PV_{\theta_e}$ is given by
\begin{equation}
    \frac{D}{Dt} \left( \frac{\vec{\omega} \cdot \nabla \theta_e} {\rho} \right) = \frac{1}{\rho^3} \nabla \theta_e \cdot \left( \nabla \rho \times \nabla p \right).
    \label{eqn:pve-evol}
    \end{equation}
Here, in the case of $PV_{\theta_e}$, the non-conservation is due to the solenoidal term. 


As a final common choice, 
one may also consider potential vorticity $PV_{\theta_l}$ defined in terms of liquid water potential temperature $\theta_l$. 
Note that $\theta_l$ is a materially conserved quantity, i.e., $D\theta_l/Dt=0$. Hence, from (\ref{eqn:pv-evol-psi}), the evolution of $PV_{\theta_l}$ is given by
\begin{equation}
    \frac{D}{Dt} \left( \frac{\vec{\omega} \cdot \nabla \theta_l} {\rho} \right) = \frac{1}{\rho^3} \nabla \theta_l \cdot \left( \nabla \rho \times \nabla p \right),
    \end{equation}
and non-conservation of $PV_{\theta_l}$ is seen to be due to the solenoidal term.


In summary,
the four moist PVs here involve four common potential temperature
variables: $\theta$, $\theta_v$, $\theta_e$, and $\theta_l$.
The discussion above serves to
illustrate the different properties of the four cases.
They are all non-conservative in different ways,
due to either sources/sinks or the solenoidal term.
For instance, $\theta_v$ has a source term due to cloud latent heating,
so $PV_{\theta_v}$ is conserved (a material invariant) in the unsaturated phase
but is not conserved in the saturated phase.
On the other hand,
the variables $\theta_e$ and $\theta_l$ are conserved,
so non-conservation of $PV_{\theta_e}$ and $PV_{\theta_l}$ 
is due to the solenoidal term only. 
Hence, it appears that there may not be a moist PV quantity 
that is a material invariant, due to clouds and phase changes.
% ...

\subsection{Moist PV is conserved over certain local volumes, even with clouds}
\label{sec:moist-pv-int}

While moist PV may be non-conservative for each fluid parcel,
it has been recently shown by \citet{kooloth2022conservation,kooloth2023hamilton}
that moist PV can be conserved when integrated over certain
local volumes
(for $PV_{\theta_e}$ or $PV_{\theta_l}$, but not $PV_{\theta}$ nor $PV_{\theta_v}$).  
Here we sketch the key ideas of these conservation principles.

As motivation for integrating over local volumes, 
start with the $PV_{\theta_e}$ evolution equation in (\ref{eqn:pve-evol}), rewritten as\footnote{
Using vector calculus identities, note that
$-\nabla \theta_e \cdot \nabla \times \left( \frac{1}{\rho} \nabla p\right) = -\nabla \theta_e \cdot \nabla \frac{1}{\rho} \times   \nabla p = \nabla \frac{1}{\rho} \cdot \nabla \theta_e \times   \nabla p = \nabla \frac{1}{\rho} \cdot \nabla \times \left( \theta_e \nabla p \right) = \nabla \cdot \left[ \frac{1}{\rho} \nabla \times \left(\theta_e \nabla p\right) \right] . $
} 
 \begin{align}
    \rho \frac{D}{Dt} \left( \frac{\vec{\omega} \cdot \nabla \theta_e} {\rho} \right) 
    = \nabla \cdot \left[ \frac{1}{\rho} \nabla \times \left(\theta_e \nabla p\right) \right].
    \label{eqn:pve-evol1}
    \end{align}
Given that a divergence appears on the right-hand side,
one might try to integrate in order to remove this divergence term.

Pursuing this direction, we integrate (\ref{eqn:pve-evol1}) 
over a material volume\footnote{Recall that, for a material volume, we have $\frac{d}{dt} \int_{V_m} dV\;\left({\vec{\omega} \cdot \nabla \theta_e} \right) = \int_{V_m} dV \;  \rho \;\frac{D}{Dt} \left(\frac{\vec{\omega} \cdot \nabla \theta_e}{\rho} \right)$.}
$V_m$ (i.e., a volume that moves with the
fluid flow) and use the divergence theorem to arrive at
\begin{align}
 \frac{d}{dt} \iiint_{V_m} dV\;\left({\vec{\omega} \cdot \nabla \theta_e} \right)  
 &=  \oiint_{S_m} d\vec{S}\; \cdot  \left( \frac{1}{\rho} \nabla \theta_e \times \nabla p \right)
 \label{eqn:pve-evol2}
 \end{align}
where $S_m$ is the material surface that bounds the material volume.
The right-hand side is still non-zero, so conservation has not yet
been demonstrated.

To simplify the right-hand side of 
(\ref{eqn:pve-evol2}),
first choose the material volume $V_m$ 
to be a distorted cylinder with base and lid given by surfaces of constant $\theta_e$ (say $\theta_e = \theta_{e1}$ and $\theta_e = \theta_{e2}$, respectively) and sides given by $q_t = q_t(\theta_e)$.
(An illustration of such a cylinder is shown below in section~\ref{sec:num-sim}.)
On the base and lid, $d\vec{S} \parallel \nabla \theta_e$  i.e., the normal to the surface is parallel to $\nabla \theta_e$ and therefore the surface integral over $S_m$ is the same as a surface integral over only the sides of the cylinder:
\begin{align}
    \oiint_{S_m} d\vec{S}\; \cdot  \left( \frac{1}{\rho} \nabla \theta_e \times \nabla p \right) &= \iint_{S_{sides}} d\vec{S}\; \cdot  \left( \frac{1}{\rho} \nabla \theta_e \times \nabla p \right) 
    \label{eqn:pve-evol3}\\
    &= - \iint_{S_{sides}} d\vec{S}\; \cdot  \nabla \times \left( g(p,\theta_e) \nabla p \right).
    \label{eqn:pve-evol4}
\end{align}
Also, to obtain the last line above, a second key observation,
from fundamentals of moist thermodynamics, is needed:
$\rho$ can be written as $\rho=\rho(p,\theta_e,q_t)$
as a function of the three moist thermodynamic quantities
$(p,\theta_e,q_t)$. Furthermore, since
$q_t=q_t(\theta_e)$ on the sides of the cylinder, 
we have $\rho=\rho(p,\theta_e,q_t(\theta_e))$
and $\rho$ is a function of $p$ and $\theta_e$ alone.
It follows that (\ref{eqn:pve-evol3}) and (\ref{eqn:pve-evol4})
are equal for a function $g(p,\theta_e)$ that satisfies
$\partial g/\partial \theta_e = 1/\rho(p,\theta_e,q_t(\theta_e))$.



To complete the derivation, by using Stokes theorem, the surface integral in (\ref{eqn:pve-evol4}) can be converted to two closed line integrals along the edges of the cylinder, $C_1$ and $C_2$, and we have
\begin{equation}
    \frac{d}{dt} \iiint_{V_m} dV\;\left({\vec{\omega} \cdot \nabla \theta_e} \right)= -\oint_{C_1} d\vec{x}\; \cdot \left( g(p,\theta_e) \nabla p \right) + \oint_{C_2} d\vec{x}\; \cdot \left( g(p,\theta_e) \nabla p \right).
\end{equation}
By noting that $\theta_e$ is a constant on both $C_1$ and $C_2$, the integrands above reduce to exact differentials which integrate to zero on the closed curves. This gives us our final result,
\begin{equation}
    \frac{d}{dt} \iiint_{V_m} \frac{\vec{\omega} \cdot \nabla \theta_e}{\rho} \;\rho \,dV = 0,
    \label{result-thetae}
\end{equation}
of conservation of $PV_{\theta_e}$ when integrated over certain
local material volumes.
\footnote{It is an open question to understand how general the material volumes could be. See \citet{kooloth2022conservation,kooloth2023hamilton} for some other known examples.}

A similar conservation law can be derived for 
$PV_{\theta_l}$ or $PV$ based on entropy or even $PV_{q_t}$
\citep{kooloth2022conservation}.
The key property shared by $\theta_e$, $\theta_l$,
$s$ (entropy), and $q_t$ is that they are all material invariants.
On the other hand, $\theta$ and $\theta_v$ are not material invariants in
the presence of phase changes and clouds, and hence
the derivation above does not hold for $PV_{\theta}$ nor  $PV_{\theta_v}$.

\section{Numerical simulations of PV conservation and nonconservation}
\label{sec:num-sim}

For numerical demonstration of the conservation laws from section~\ref{sec:background}, we will set aside the compressible
setting that includes acoustic/sound waves and use the simpler
setting of the Boussinesq approximation. 
The governing equations under the Boussinesq approximation
are described in the Supporting Information,
and they are similar to equations of moist Boussinesq dynamics that have been used in other studies
\citep[e.g.,][]{kuo1961convection,bretherton1987theory,grabowski1993cloud,pauluis2010idealized,stechmann2010multiscale,stechmann2014multiscale,hmss13,marsico2019energy}.
The Boussinesq case admits 
statements of PV conservation and non-conservation
\citep{kooloth2023hamilton}
that are analogous to the compressible case
from section~\ref{sec:background}.
A summary is as follows.

As a particular moist PV quantity for illustration, we use
$PV_u$ which is based on the total buoyancy $b_u$ in the unsaturated phase:
\begin{align}
PV_u = {\Vec{\omega}}\cdot \nabla b_u.
\end{align}
The evolution of $PV_u$ is then given by
\begin{align}
\frac{D}{Dt} PV_u=
     \frac{D}{Dt} \left( {\vec{\omega} \cdot \nabla b_u}  \right) =   \nabla b_u \cdot (\nabla \times b' \hat{z}).
    \label{eqn:pvu-evol}
\end{align}
The buoyancy $b'$ 
depends on the phase and is nonconservative. Consequently, it appears as a source term in the evolution of $PV_u$, which is then also nonconservative. 

As a special case, though, note that the right-hand-side goes to zero in the unsaturated phase since $\nabla b_u \cdot (\nabla \times b_u' H_u \hat{z}) = 0,$ and therefore 
\begin{align}
\frac{D}{Dt} PV_u=
     \frac{D}{Dt} \left( {\vec{\omega} \cdot \nabla b_u}  \right) =  0
     \quad\mbox{if unsaturated,}
    \label{eqn:pvu-evol-unsat}
\end{align}
so that $PV_u$ is pointwise conserved for any parcels that are not inside a cloud. 


For a general scenario involving phase changes,
following similar steps as presented for the compressible case in section \ref{sec:moist-pv-int}, a parcel-integrated $PV_u$ conservation principle can be obtained:
\begin{equation}
    \frac{d}{dt} \iiint_{V_m} \vec{\omega} \cdot \nabla b_u  \,dV = 0,
    \label{eqn:pvu-evol-int}
\end{equation}
which follows from integrating (\ref{eqn:pvu-evol}) over a material volume $V_m$ 
that is a distorted cylinder
whose base and lid are given by $b_u = const.$ and the sides are given by $b_s=const.$,
where $b_s$ is the total buoyancy in the saturated phase.
This $PV_u$ conservation statement can also be shown to be valid for a material volume enclosed by isosurfaces of the more physically relevant quantities, $\theta_e$ and $q_t$. The main ideas of the derivation are the same as in the compressible case; the interested reader can refer to supporting information for the detailed derivation.

Parcel-integrated conservation principles can also be derived for many other potential vorticity quantities for the Boussinesq system. 
One such quantity is potential vorticity $PV_s ={\Vec{\omega}}\cdot \nabla b_s$ based on the total saturated buoyancy;
two others are $PV_e = {\Vec{\omega}}\cdot \nabla \theta_e$ based on equivalent potential temperature $\theta_e$, and $PV_l = {\Vec{\omega}}\cdot \nabla \theta_l$ based on liquid water potential temperature $\theta_l$. 
Additionally, it can be shown that $PV_s$
is materially conserved in the saturated phase using a similar reasoning as for $PV_u$
in the unsaturated phase
\citep{kooloth2023hamilton}.

\begin{figure}[h!]
    \centering
    \includegraphics[width=\textwidth]{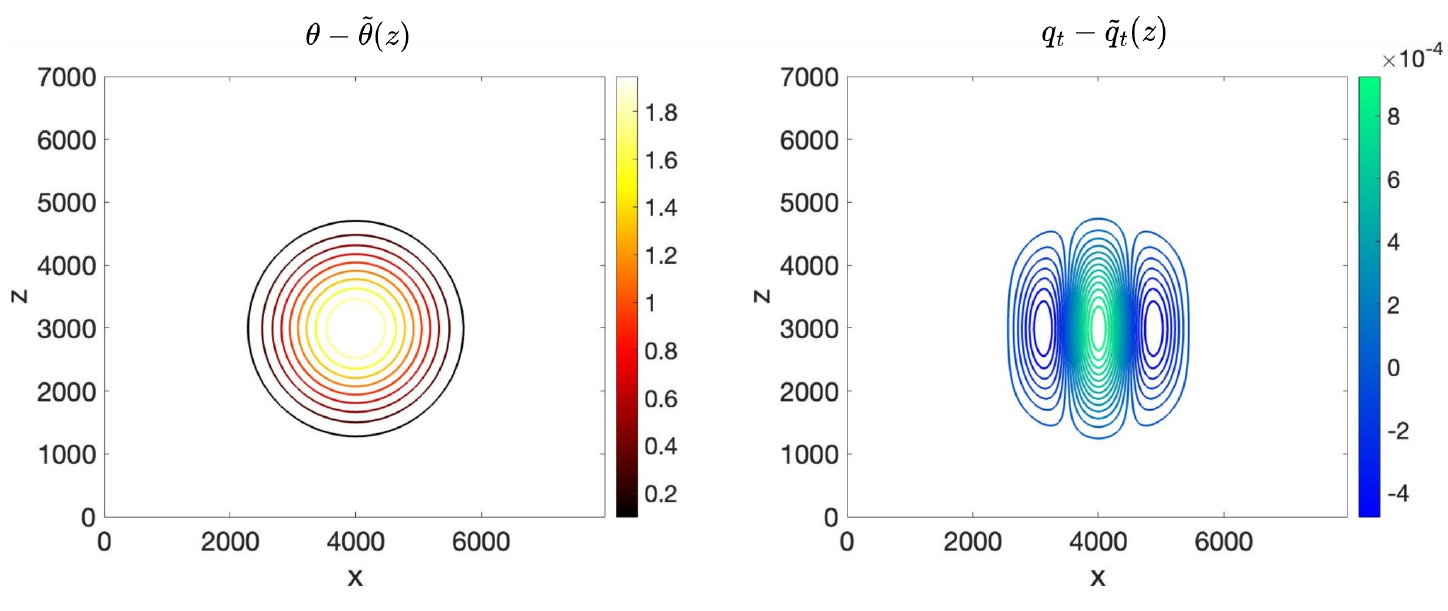}
    \caption{Initial conditions for the 3D rising thermal. Plots are shown in the $(x,z)$ plane with fixed $y=0.5L_y = 3500 \; {\rm m}$. Anomalies of potential temperature $\theta$ (left) and total water specific humidity $q_t$ (right), defined as anomalies from a horizontally uniform background state, $\tilde{\theta}(z)$ and $\tilde{q}_t(z)$. The units for $x$ and $z$ are meters, and the units for $\theta$ and $q_t$ are K and kg/kg respectively.}
    \label{fig:initial}
\end{figure}

\begin{figure}[h!]
    \centering
    \includegraphics[width=1\textwidth]{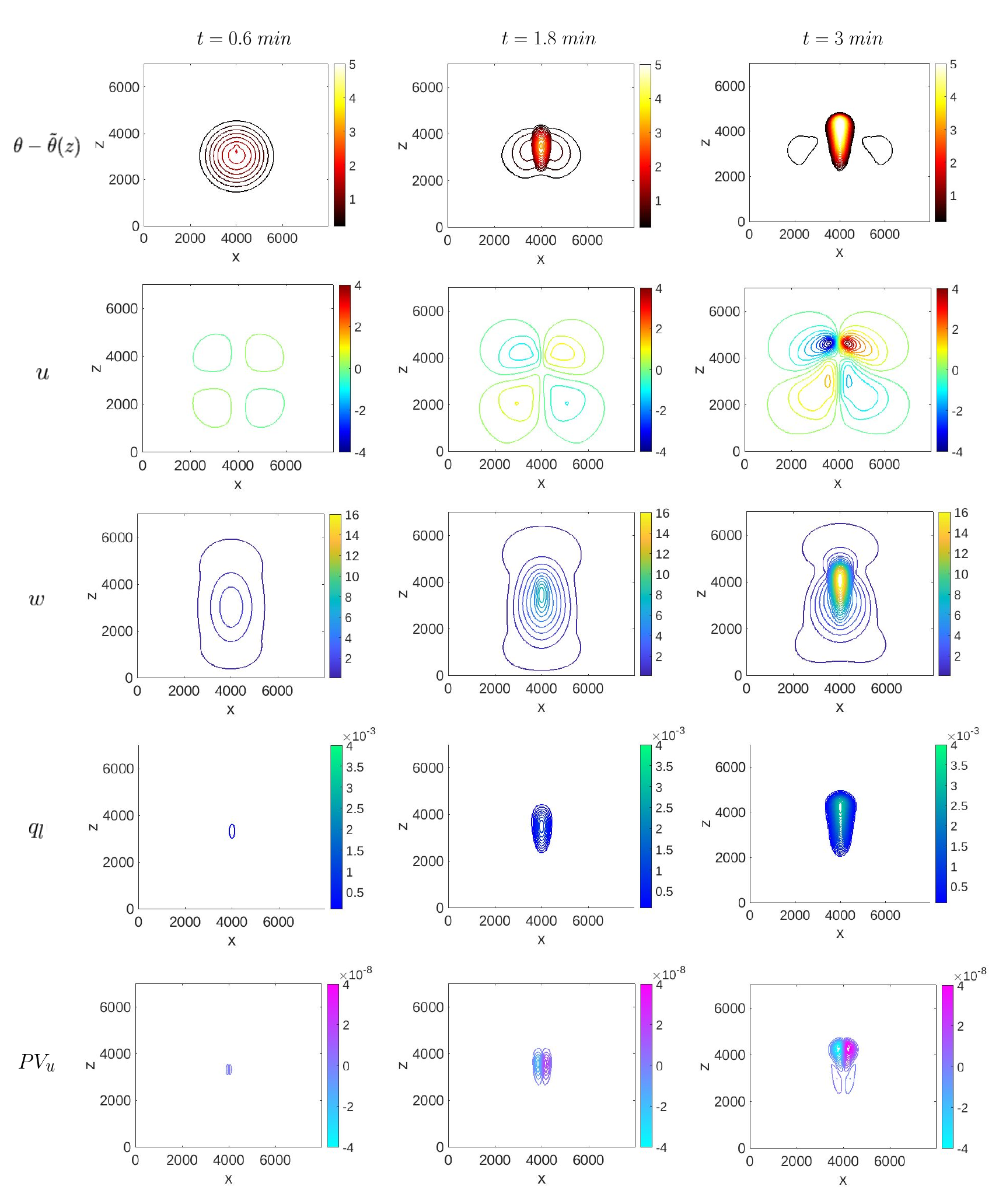}
    \caption{Snapshots of the warm bubble in the plane with fixed $y=0.5L_y=3500$ m.
    Rows show $\theta - \tilde{\theta}$, $q_l$, $u$, $w$ and $PV_u$ and columns show times $t=0.6$, $1.8$ and $3$ minutes. The units for $x$ and $z$ are meters, and the units for $\theta$, velocities ($u,\; v,\; w$), $q_l$ and $PV_u$ are K, m/s, kg/kg and s$^{-3}$ respectively.}
    \label{fig:snapshot}
\end{figure}

The three-dimensional (3D) numerical simulations in this study are performed using the code of \citet{hmss13}. The channel domain 
is periodic in the $x$ and $y$ directions, and assumes a rigid top and bottom.  
A third-order Runge-Kutta scheme with 
 adaptive time-stepping (CFL $= 0.9$) is used for time integration. Spatial discretization is based on pseudo-spectral decomposition using Fourier modes in the horizontal directions, and 2nd-order centered differences on a staggered grid in the vertical direction. 
 In our study, the domain size $L_x\times L_y\times H$ is
 $8000 \;\mbox{m} \times 8000 \;\mbox{m} \times 7000 \;\mbox{m}$ and the number of grid points is $256 \times 256 \times 400$,
 corresponding to horizontal and vertical grid spacings of $31.5$ m and $17.5$ m, respectively.

The case study for illustration is the well-known case
of a rising moist thermal, and the basic aspects of the simulation
are as follows.
A horizontal slice through the initial, unsaturated, spherical perturbation is shown in Figure \ref{fig:initial} (at $y=3500 \; {\rm m}$).  Evolution of the perturbation in an $(x,z)$ plane is shown in Figure \ref{fig:snapshot}.  For $t>0$, the warm bubble rises due to its buoyancy and a non-zero velocity field develops. 
In the plane at $y=3500$ m, the vapor starts to condense near the center of the plane at $t \approx 1.8$ min, contributing to the formation of a 3D cloud (Figures \ref{fig:snapshot} and \ref{fig:material_vol_curve}). 
The size of the cloud grows as more fluid parcels change phase (see Figure \ref{fig:snapshot} at $t=3$ min and Figure \ref{fig:material_vol_curve} at $t=1.2$ min).\subsection{Local-volume integrated $PV_u$ conservation}
\label{subsec:volume-conservation-numerical}

\begin{figure}
    \centering
    \includegraphics[width=1\textwidth]{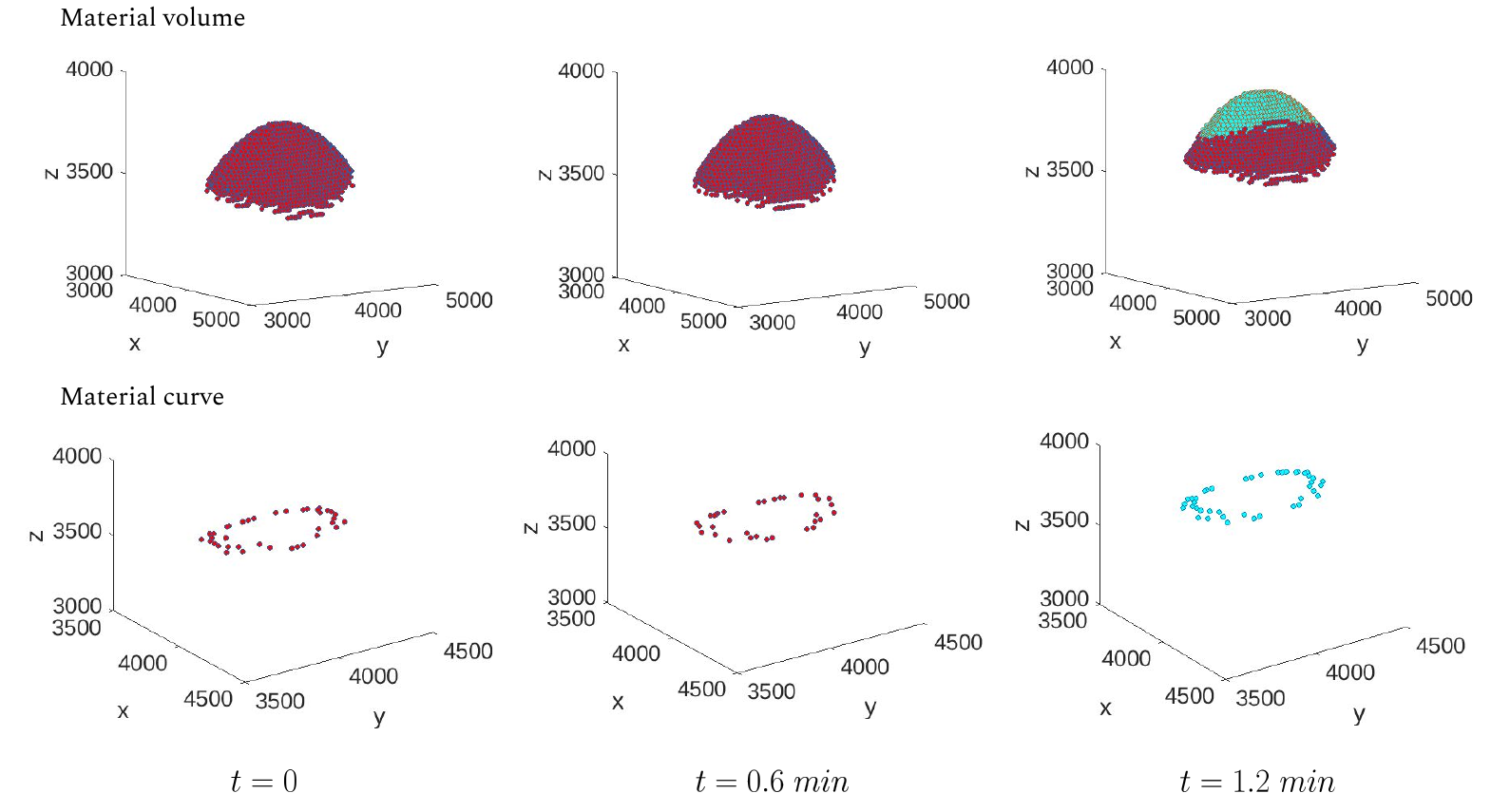}
    \caption{Illustration of a material volume and material curve, at times $t=0$, 0.6, and 1.2 min. (Top row) Evolution of a material volume enclosed between $\theta_e> 326$ K and $ 2.0 \times 10^{-3} < q_t < 5.75 \times 10^{-3}$ (in kg/kg). The red and the cyan dots represent the unsaturated and saturated parcels respectively. (Bottom row)  Evolution of a closed material curve of $\theta_e \approx 326.48$ K and $q_t \approx 5.66 \times 10^{-3}$ kg/kg. }
    \label{fig:material_vol_curve}
\end{figure}

In order to verify 
volume-integrated $PV_u$ conservation, 
a material volume is identified and tracked over time in the rising bubble simulation described above. The material volume consists of roughly 5000 grid cells and is specified by certain level surfaces for $\theta_e$ 
and  $q_t$, such that their intersection encloses a moving material volume. 
As shown in Figure \ref{fig:material_vol_curve}, at the earlier times of $t = 0$ min and $t=0.6$ min, the fluid parcels in the material volume are unsaturated. By the later time of $t=1.2$ min, however, a cloud of liquid water has developed in upper levels of the volume, and $47\%$ of the fluid parcels within the material volume have 
undergone a change of phase in their water content, from water vapor to liquid water.
Concurrently, 
within the cloud,
the local values of $PV_u$ are evolving (see Figure \ref{fig:snapshot}).

On the other hand, we can numerically check for local volume-integrated $PV_u$ by 
computing 
\begin{equation}
\label{def:ivp}
    IPV = \iiint_{V_m} dV \; PV_u,
\end{equation}
where $V_m$ is the specified material volume.
Noting that the initial velocity field is zero everywhere in the domain, then the $IPV$ should start and stay at the value zero to machine precision.
By monitoring \eqref{def:ivp} in time, we found that the $IVP$ within the material volume remains $O(10^{-16})$ for the entire simulation,
verifying the conservation statement 
for parcel-integrated PV.





\subsection{Material $PV_u$ conservation and non-conservation}

In this section, we investigate the material conservation of $PV_u$.
It is shown that $PV_u$ is materially conserved prior to the time when fluid parcels undergo phase change from vapor to liquid, but not 
for later times after the cloud has formed.  We identify and track a closed material curve within the material volume, 
which are both rising along with the bubble, as shown on the second row of Figure \ref{fig:material_vol_curve}. The material curve is initially unsaturated, but by $t=1.2$ min, all the parcels composing the material curve have undergone a change of phase. 

To quantify material conservation and non-conservation, we measured the maximum of the absolute value of $PV_u$ within the material volume (Figure \ref{fig:material_vol_curve}).
At representative early time $t=0.6,$ before a significant number of parcels have experienced a change of water phase,
the max absolute value of $PV_u$ within the entire material volume is $6.8 \times 10^{-11}$ s $^{-3}$.  In comparison, at later time $t=1.2$ min, by which time a robust cloud has formed, the max absolute value of $PV_u$ within the volume has increased by roughly 3 orders of magnitude, obtaining the value $6.8 \times 10^{-8}$ s $^{-3}$. 

\begin{figure}
    \centering
\includegraphics[width=0.6\textwidth]{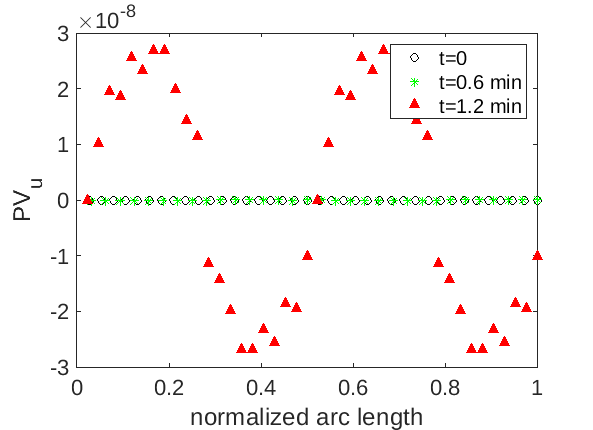}
    \caption{Demonstration of parcel-wise non-conservation of $PV_u$ (s$^{-1}$), due to phase change between $t=0.6$ and 1.2 min. Evolution of $PV_u$ on the closed material curve of $\theta_e^{tot} \approx 326.48$ K and $q_t^{tot} \approx 0.00566$ kg/kg. The points on the material curve are ordered using the arc length coordinate.}
    \label{fig:pv_curve}
\end{figure}

As a graphical illustration, Figure \ref{fig:pv_curve}
shows the $PV_u$ values along the selected material curve.
The figure shows $PV_u$ associated with each grid point along the curve, at three different times $t=0$, $0.6$, $1.2$ minutes. At $t=0$ and $t=0.6$ min, when the material curve is unsaturated, the $PV_u$ values are close to zero. At $t=1.2$ min, some $PV_u$ values associated with the material curve are of the order of $10^{-8}$ s$^{-1}$, demonstrating that $PV_u$ does not remain conserved along the material curve once the parcels undergo change of water phase.



\section{Conclusions}
\label{sec:conclusions}

In this work, a main goal was to compare and contrast the different
moist PV conservation statements and non-conservation statements,
in light of the rich variety of possibilities 
in the literature, including recent developments
\citep[e.g.][]{bennetts1979conditional,emanuel1979inertial,schubert2001potential,marquet14definition,kooloth2022conservation,kooloth2023hamilton}.

As a summary, Table \ref{tab:PV_summary}
lists the different formulations of PV considered here and their conservation laws. We note that in a moist flow with phase changes, there are no material invariant PVs; the strongest conservation principle realizable in this setting is an integrated PV invariance for certain material volumes. In both the fully compressible and Boussinesq cases, PVs based on $\theta_e$ and $\theta_l$ possess a material-volume-integrated conservation principle, even with phase transitions. In the moist Boussinesq setting, $PV_u$ and $PV_s$, based on the unsaturated and saturated buoyancy, respectively, also have an integrated PV invariance principle within certain material volumes even with phase changes. Additionally, as special cases, $PV_u$ and $PV_s$ remain materially invariant in the unsaturated and saturated phases respectively.

A more complete set of desirable properties is often associated with PV,
including conservation, inversion, and balance/slow variation.
Table~\ref{tab:PV_summary} also summarizes these other properties.
As one note about inversion and balance/slow variation,
since $PV_{\theta_v}$ is not slowly varying, any inversion with
$PV_{\theta_v}$ will not completely remove the fast wave contributions
in the presence of phase changes \citep{wetzel2020potential}.
The moist PV quantities that are defined in terms of conserved thermodynamic
variables (e.g., $\theta_e$ or $\theta_l$) are slowly evolving,
and they are associated with a PV-and-M inversion principle (see the discussion of inversion in section~\ref{sec:intro}).

One application of parcel-integrated PV conservation laws
is for diagnosing diabatic processes in the atmosphere or ocean
(see discussion in section~\ref{sec:intro}).
In the past, (parcel-wise) PV non-conservation was often used to indicate
cloud latent heating, as a leading diabatic process.
In contrast, for the new conservation laws of PV as a
parcel-integrated invariant, the conservation law holds
even in the presence of phase changes and cloud latent heating;
consequently, any non-conservation must be due to other diabatic processes,
such as friction/viscosity, radiative cooling/heating, or precipitation.
Therefore, the parcel-integrated conservation laws can
potentially provide new information about other diabatic processes
in diagnostic studies.

\begin{table}[bt]
\caption{Summary of various PV formulations and their conservation, inversion, and balance/slow variation properties.}
\label{tab:PV_summary}
\begin{tabular}{ccccc}
\headrow
\thead{PV definition} & \thead{Material invariant} & \thead{Material-volume-integrated invariant} & \thead{Inversion$^*$} & \thead{Slowly varying$^{**}$} \\
\hiderowcolors
\thead{Dry} \\
$PV_{\theta} = \frac{\vec{\omega} \cdot \nabla \theta} {\rho}$ & everywhere &  all $V_m$ & PV & Yes\\
\thead{Moist} \\
$PV_{\theta_v} = \frac{\vec{\omega} \cdot \nabla \theta_v} {\rho}$ & unsaturated phase &  all unsaturated $V_m$ & None & No\\
$PV_{\theta_e} =\frac{\vec{\omega} \cdot \nabla \theta_e} {\rho}$ & nowhere & certain $V_m$, even with phase changes & PV-and-M & Yes \\
$PV_{\theta_l} =\frac{\vec{\omega} \cdot \nabla \theta_l} {\rho}$ & nowhere & certain $V_m$, even with phase changes  & PV-and-M & Yes  \\
\thead{Moist Boussinesq}  \\
$PV_{u} ={\vec{\omega} \cdot \nabla b_u} $ & unsaturated phase & certain $V_m$, even with phase changes,  & PV-and-M & Yes \\ & & and also all unsaturated $V_m$ \\
$PV_{s} ={\vec{\omega} \cdot \nabla b_s} $ & saturated phase &  certain $V_m$, even with phase changes,  & PV-and-M & Yes \\ & & and also all saturated $V_m$ \\
$PV_{e} ={\vec{\omega} \cdot \nabla \theta_e} $ & nowhere &  certain $V_m$, even with phase changes  & PV-and-M & Yes \\
$PV_{l} ={\vec{\omega} \cdot \nabla \theta_l} $ & nowhere & certain $V_m$, even with phase changes  & PV-and-M & Yes \\
\hline  
\end{tabular}
\begin{tablenotes}
\item $^*$ to recover the balanced or slowly evolving component(s) of the system 
\item $^{**}$ with phase changes in the moist cases.
\item $V_m$ refers to a material volume.
\end{tablenotes}
\end{table}

\section*{acknowledgements}
The authors thank Gerardo Hernandez--Duenas for providing the
computer code used for the simulations.
L.M. Smith and S.N. Stechmann gratefully
acknowledge support from US NSF grant DMS-1907667.


\printendnotes

\bibliography{references}



\end{document}


\title{Supporting Information for: \\Non-conservation and conservation for different formulations of moist potential vorticity}
\author{Parvathi Kooloth, Leslie M. Smith, Samuel N. Stechmann 
\\
November 10, 2023}

\maketitle

\section*{Appendix S1: Potential vorticity for Boussinesq systems}
\label{sec:bouss}

For numerical demonstration of the conservation laws from the main text, we set aside the compressible
setting that includes acoustic/sound waves and use the simpler
setting of the Boussinesq approximation. 
In this setting, we describe the governing equations
 and a proof of PV conservation and non-conservation.


\subsection*{Governing Equations}
\label{sec:gov-eqn}




Under the Boussinesq approximation, the 
governing equations of the moist compressible atmosphere in 
(1) and (3) of the main text are changed to
\begin{subequations}
\label{boussinesq-system}
\begin{align}
    \frac{D\vec{u}}{Dt} &= -\frac{1}{\rho_0}\nabla p^\prime+ b' \hat{z}, \label{eqn:u-evol} \\
    \frac{D\theta_e}{Dt} &= 0, 
    \label{eqn:thetae-evol} \\
    \frac{Dq_t}{Dt} &= 0,
    \label{eqn:qt-evol} \\
    \nabla \cdot \vec{u} &=0,
    \label{eqn:incomp}
\end{align}
\end{subequations}
where $\vec{x} = (x,y,z)$ is the position vector, $\hat{z}$ is the unit vector in the $z$-direction, $\rho_0$ is the constant background density,
$\vec{u} = (u,v,w)$ is the velocity, and 
$D/Dt=\partial/\partial t+\vec{u}\cdot\nabla$ is the material derivative.
For similar equations for moist Boussinesq dynamics, see, e.g., \cite{kuo1961convection,bretherton1987theory,grabowski1993cloud,pauluis2010idealized,stechmann2010multiscale,stechmann2014multiscale,hmss13,kooloth2023hamilton,marsico2019energy}.
Any thermodynamic variables could be decomposed into a background function of height and an anomalous part, e.g.,  the equivalent potential temperature $\theta_e(\Vec{x},t)=
\Tilde{\theta}_e(z) + \theta_e^\prime(\Vec{x},t),$
where $\Tilde{\theta}_e(z)$ is prescribed. For liquid water $q_l$, the background state is zero.  Thus $p'$ denotes the pressure anomaly and $b'$ represents the buoyancy. 
The total water $q_t$ is the sum of water vapor $q_v$ and liquid water $q_l$,
\begin{equation}
    q_t=q_v+q_l,
    \label{eqn:qt-def}
\end{equation} 
and the equivalent potential temperature $\theta_e$ is related to the potential temperature $\theta$ by
 \begin{equation}
      \theta_e = \theta + (L/c_p)q_v,
      \label{eqn:thetae-def}
 \end{equation}
where the latent heat $L$ 
and specific heat $c_p$ are constants with values $L \approx 2.5 \times 10^6$ J kg$^{-1}$ and $c_p \approx 10^3$ J kg$^{-1}$ K $^{-1}$.
The buoyancy $b'$ may be expressed as 
\begin{align}
     b' &= g \left( \frac{(\theta-\tilde{\theta}(z))}{\theta_0} + R_{vd}(q_v- \tilde{q}_v(z)) -q_l  \right),
     \label{eqn:b-def-theta}
\end{align}
where 
$R_{vd}$ is the ratio of gas constants of dry and moist air. 

To fully specify (\ref{boussinesq-system}) in closed form,
it is necessary to describe
the buoyancy $b'$ in terms of the prognostic variables
$\theta_e$ and $q_t$ from (\ref{boussinesq-system}).
Hence, we must describe a transformation of variables from
$(\theta_e,q_t)$ to $(\theta,q_v,q_l)$,
as the inverse of the transformation in (\ref{eqn:qt-def}) and (\ref{eqn:thetae-def}).
To do so, we use different formulas for the transformation,
depending on the two possible states of the water content, saturated or unsaturated, and these two states are separated by a phase boundary. 
The state is 
determined by comparing the magnitude of the total water specific humidity $q_t$ to the magnitude of a saturation specific humidity $q_{vs}$, which acts as a threshold.
For simplicity, $q_{vs}$ is approximated by a function of height only such that $q_{vs} \approx q_{vs}(z)$ (see \cite{hmss13} and equation \eqref{eqn-qvs} below).
In unsaturated flow regions with  $q_t<q_{vs}$, the water is all in the form of vapor 
such that $q_t=q_v,$ and $q_l=0.$
On the other hand, in saturated regions with 
$q_t \geq q_{vs},$
then the water vapor is exactly at saturation, and the remaining water is liquid, such that 
$q_t=q_{vs} + q_l$.
Therefore, the formulas for $q_v$ and $q_l$ may be written
\begin{align}
    q_v &= \min(q_t,q_{vs}),
    \label{eqn:qv-def} 
\\
    q_l &= \max(0,q_t-q_{vs}).
    \label{eqn:ql-def}
\end{align}
Similarly, to recover $\theta$ from $\theta_e$ and $q_t$,
we write (\ref{eqn:thetae-def}) in the unsaturated and saturated phases as
$\theta_e=\theta+(L/c_p)q_t$ and $\theta_e=\theta+(L/c_p)q_{vs}$,
respectively, and rearrange and combine to arrive at
\begin{equation}
    \theta=\theta_e-(L/c_p)\min (q_t,q_{vs}).
    \label{eqn:theta-def-si}
\end{equation}
Together, (\ref{eqn:qv-def})--(\ref{eqn:theta-def-si})
show how to write $(\theta,q_v,q_l)$ in terms of 
$(\theta_e,q_t)$, which can be used in (\ref{eqn:b-def-theta})
to write the buoyancy in terms of the prognostic variables
$\theta_e$ and $q_t$.

Furthermore, the presence of phase changes can be made explicit in another way by re-writing the buoyancy $b^\prime$ from (\ref{eqn:b-def-theta}) as
\begin{align}
    b' = b_u' H_u + b_s' H_s,
\end{align}
where the subscript $u \;(s)$ denotes the unsaturated (saturated) state,
$H_u, H_s$ are Heaviside functions and $b_u^\prime, b_s^\prime$ are the expressions for the buoyancies in the different states.  More specifically,
the Heavside function $H_u = H(q_{vs} - q_t)$ has the two values $H_u = 1$ for $q_t < q_{vs}$ and $H_u = 0$ for $q_t \geq q_{vs}$, with $H_s = 1- H_u$. To find the expressions for the buoyancies $b_u'$ and $b_s'$ in terms of the conserved quantities, $\theta_e$ and $q_t$, we insert (\ref{eqn:qv-def})--(\ref{eqn:theta-def-si})
into the $b'$ formula in (\ref{eqn:b-def-theta}).
The resulting formulas in the unsaturated and saturated phases
are the formulas for $b_u'$ and $b_s'$:
\begin{equation}
 b_u^\prime = g \left( \frac{\theta_e-\tilde{\theta}_e(z)}{\theta_0} + \left(R_{vd} -\frac{L_v}{c_p \theta_0} \right) (q_t- \tilde{q}_t(z)) \right), 
 \quad
 b_s^\prime =g \left( \frac{\theta_e-\tilde{\theta}_e(z)}{\theta_0} + \left(R_{vd} -\frac{L_v}{c_p \theta_0} + 1\right) q_{vs}(z)- (q_t- \tilde{q}_t(z)) \right).
 \label{def:bubs}
\end{equation}
The unsaturated buoyancy $b_u^\prime$ will be used below
in defining a potential vorticity variable for illustrating
PV conservation and non-conservation.

\subsection*{Potential Vorticity $PV_u$}
\label{sec:pvu}

For our numerical study, it is convenient to introduce the moist potential vorticity $PV_u$ based on buoyancy in unsaturated flow regions.  As will be discussed, this choice allows as to compare material non-conservation (following individual fluid parcels) to 
volume-integrated conservation (for parcels within a moving material volume).

The moist potential vorticity $PV_u$ is based on the total buoyancy $b_u$ in the unsaturated phase, such that
\begin{align}
PV_u = {\Vec{\omega}}\cdot \nabla b_u,
\end{align}
where 
\begin{equation}
b_u = g\left( \frac{\theta_e}{\theta_0} + \left(R_{vd} -\frac{L_v}{c_p \theta_0}\right) q_t\right) = b_u^\prime + N_u^2 z,
\label{eqn:bu-def-si}
\end{equation} 
and where $N_u$ is the buoyancy frequency in the unsaturated regions of the flow
\begin{equation}
N_u = \biggl [ g \frac{d}{d z} 
\biggl (\frac{\tilde{\theta}_e}{\theta_0} + \left(R_{vd} -\frac{L_v}{c_p \theta_0}\right)\tilde{q}_t\biggr )
\biggr ]^{1/2}.
\end{equation} 
Similarly, one could define a total buoyancy in the saturated phase as
$b_s = b_s^\prime + N_s^2 z$, where $b_s^\prime$ is given by \eqref{def:bubs} and
$N_s^2$ is the associated saturated buoyancy frequency \citep{smith2017precipitating}.

\subsection*{Material Volume-integrated Conservation of $PV_u$}

Derivations of PV conservation laws for the moist Boussinesq
equations are as follows.
See also \cite{kooloth2023hamilton}.

\paragraph{Parcels bounded by surfaces of constant $b_u$ and $b_s$}

We start with the governing equation for PV,
\begin{align}
    \rho \frac{D}{Dt} \left( {\vec{\omega} \cdot \nabla b_u}  \right) &=  \nabla b_u \cdot \nabla \times \left( b' \nabla z\right).
\end{align}
Now we integrate over a material volume that is a distorted cylinder with base and lid given by surfaces of constant $b_u$ (say $b_u = b_{u1}$ and $b_u=b_{u2}$) and sides parametrized as $b_s = b_s(b_u)$:
\begin{align}
    \frac{d}{dt} \iiint_{V_m} dV\;\left({\vec{\omega} \cdot \nabla b_u} \right) &= \frac{d}{dt} \iiint_{V_a} dV_a J \left({\vec{\omega} \cdot \nabla b_u} \right)\\
    & = \iiint_{V_a} dV_a \;\frac{\partial}{\partial t} \left({\vec{\omega} \cdot \nabla b_u} \right)\\
    &= \iiint_{V_m} dV \;  \rho \;\frac{D}{Dt} \left({\vec{\omega} \cdot \nabla b_u} \right).
\end{align}
The second line results from knowing $J = \frac{\partial (\vec{x})}{\partial (\vec{a})} = \frac{1}{\rho}$ by appropriately choosing $\vec{a}$ is in \cite{salmon1998lectures}. And in the moist Boussinesq case, we have $\rho = 1$. Now, using the divergence theorem, since
$\nabla b_u \cdot \nabla \times \left( b' \nabla z\right) = \nabla \cdot \left(b_u\nabla\times \left(b' \nabla z\right)\right),$
 we have,
\begin{align}
    \frac{d}{dt} \iiint_{V_m} dV\;\left({\vec{\omega} \cdot \nabla b_u} \right) &= \iiint_{V_m} dV\;  \nabla \cdot \left( b_u \nabla \times \left(b' \nabla z \right) \right),\\
    &= \oiint_{S_m} d\vec{S}\; \cdot  b_u \nabla \times \left( b' \nabla p\right) =  \oiint_{S_m} d\vec{S}\; \cdot  \left( \nabla \times \left( b' b_u \nabla z\right) - \left( b' \nabla b_u \times \nabla z \right)\right), \\
    &=  -\oiint_{S_m} d\vec{S}\; \cdot  \left( b' \nabla b_u \times \nabla z \right) \label{sides}.
\end{align}
On the base and lid, we have $d\vec{S} \parallel \nabla b_u$  i.e., the normal to the surface is parallel to $\nabla b_u$;
as a result, the integrand is zero on the base and the lid, and the integral over the entire surface $S_m$ is the same 
as the integral over only the sides. 
Also, if $b_s = b_s(b_u)$ on the sides, 
$$-\iint_{S_{sides}} d\vec{S}\; \cdot  \left( b' \nabla b_u \times \nabla z \right)= -\iint_{S_{sides}} d\vec{S}\; \cdot  \left( f(z,b_u) \nabla b_u \times \nabla z \right) = -\iint_{S_{sides}} d\vec{S}\; \cdot  \nabla \times \left( g(z,b_u) \nabla z \right).
$$where $\partial g(b_u,z)/ \partial b_u = f(b_u,z)$. Now using Stokes' theorem, we have
\begin{align}
    -\iint_{S_{sides}} d\vec{S}\; \cdot  \left( b \nabla b_u \times \nabla z \right)= -\int_{C_1} d\vec{x}\; \cdot \left( g(z,b_u) \nabla z \right) + \int_{C_2} d\vec{x}\; \cdot \left( g(z,b_u) \nabla z \right) 
\end{align}
where the curves $C_1$ and $C_2$ are the two edges of the 
distorted cylinder.
Finally, note that each of the two curves $C_1$ and $C_2$
lies on a surface of constant $b_u$; therefore,
letting $C$ represent either $C_1$ or $C_2$,
the two line integrals above can be evaluated as
\begin{equation}
    \int_{C} d\vec{x}\; \cdot \left( g(z) \nabla z \right)
    = \int_{C} d\vec{x}\; \cdot \nabla G(z) =  \int_{C} dG(z) = 0,
\end{equation}
where $G$ is an antiderivative of $g$.
This completes the derivation and shows that
$PV_u$, when integrated over a certain material volume $V_m$,
is conserved.

One can also derive the analogous conservation statement for $PV_s$ in a local volume with base and lid as surfaces of constant $b_s$ (say $b_s = b_{s1}$ and $b_s=b_{s2}$) and sides given by $b_u = b_u (b_s)$, by using the same strategy.

\paragraph{Parcels bounded by surfaces of constant $\theta_e$ and $q_t$}

The derivation above can be modified to work for a distorted cylinder with base and lid given by surfaces of constant $\theta_e$ (say $\theta_e = \theta_{e1}$ and $\theta_e = \theta_{e2}$) and sides parameterized as $q_t = q_t(\theta_e)$. Using $\theta_e (= \theta + \frac{L_v}{c_p}q_v)$, $b_u$ can be expressed as a function of $\theta_e$ and $q_t$ . This gives us
\begin{equation}
    \nabla b_u = g\left( \frac{1}{\theta_0} \nabla \theta_e + \Big(R_{vd} - \frac{L_v}{c_p} \Big) \nabla q_t \right).
\end{equation}
Therefore we have,
\begin{align}
    \frac{d}{dt} \iiint_{V_m} dV\;\left({\vec{\omega} \cdot \nabla b_u} \right) &=  -\oiint_{S_m} d\vec{S}\; \cdot  \left( b' g \left( \frac{1}{\theta_0} \nabla \theta_e + \Big(R_{vd} - \frac{L_v}{c_p} \Big) \nabla q_t \right) \times \nabla z \right) \\
    &=  -\oiint_{S_m} d\vec{S}\; \cdot  \left( b' g \left( \frac{1}{\theta_0} \nabla \theta_e \right)\times \nabla z \right)  -\oiint_{S_m} d\vec{S}\; \cdot  \left( b' g \left(  \Big(R_{vd} - \frac{L_v}{c_p} \Big) \nabla q_t\right) \times \nabla z \right)  \\
    &= I_1 + I_2 \label{sides}.
\end{align}
For ease of exposition, the first surface integral term with $\theta_e$ above will be referred to as $I_1$ and the latter integral in terms of $q_t$ will be called $I_2$. On the base and lid, $d\vec{S} \parallel \nabla \theta_e$  i.e., the normal to the surface is parallel to $\nabla \theta_e$. So, $I_1$ in the equation above receives zero contribution from integrating over the base and the lid of $S_m$. Similarly, $I_2$ receives zero contribution from integrating over the sides of $S_m$ where $q_t=const.$ On the sides, $b'$ can be written as a function of $\theta_e$ and $z$ only, therefore
\begin{align}
    I_1 = -\iint_{S_{sides}} d\vec{S}\; \cdot  \left( b' \frac{g}{\theta_0}\nabla \theta_e \times \nabla z \right)= -\iint_{S_{sides}} d\vec{S}\; \cdot  \left( f(z,\theta_e) \nabla \theta_e \times \nabla z \right) = -\iint_{S_{sides}} d\vec{S}\; \cdot  \nabla \times \left( h(z,\theta_e) \nabla z \right)
\end{align}
where $\partial h(\theta_e,z)/ \partial \theta_e = f(\theta_e,z)$ and $C_1$ and $C_2$ are the top and bottom edges of the cylinder. Now using Stokes' theorem, we have,
\begin{align}
    I_1= -\int_{C_1} d\vec{x}\; \cdot \left( g(z,\theta_e) \nabla z \right) - \int_{C_2} d\vec{x}\; \cdot \left( g(z,\theta_e) \nabla z \right),
\end{align}
where $C_1$ and $C_2$ are the top and bottom edges of the cylinder.
Note that on $C_1$ and $C_2$, both $\theta_e$ and $q_t$ are constants and so,
\begin{align}
I_1
    &= -\int_{C} d\vec{x}\; \cdot \left( g(z) \nabla z \right), \\
    &= -\int_{C} d\vec{x}\; \cdot \nabla G(z) =  \int_{C} dG(z) = 0.
\end{align}
where $C = C_1 \cup C_2$ and $G$ is an antiderivative of $g$. Similarly, for $I_2$, on the base and the lid where $\theta_e$ is constant, $b'$ can be written as a function of only $q_t$ and $z$. Following the same steps as above, we have
\begin{align}
    I_2 = -\iint_{S_{top/base}} d\vec{S}\; \cdot  \left( b'g  \left( \Big(R_{vd} - \frac{L_v}{c_p} \Big) \nabla q_t \right) \times \nabla z \right)= -\int_{C_1} d\vec{x}\; \cdot \left( h'(z,q_t) \nabla z \right) - \int_{C_2} d\vec{x}\; \cdot \left( h'(z,q_t) \nabla z \right) 
\end{align}
and
\begin{align}
   I_2
    &= \int_{C} d\vec{x}\; \cdot \left( h'(z) \nabla z \right), \\
    &= \int_{C} d\vec{x}\; \cdot \nabla F(z) =  \int_{C} dF(z) = 0.
\end{align} And therefore
\begin{align}
    \frac{d}{dt} \iiint_{V_m} dV\;\left({\vec{\omega} \cdot \nabla b_u} \right) &=  0,
\end{align}
so that volume-integrated $PV_u$ is conserved for
material volumes $V_m$ that are distorted cylinders with
base and lid given by surfaces of constant $\theta_e$ (say $\theta_e = \theta_{e1}$ and $\theta_e = \theta_{e2}$) and sides parameterized as $q_t = q_t(\theta_e)$.

\section*{Appendix S2: Setup of Numerical Simulations}
\label{sec:num-setup}

The 3D simulations in this study are performed using the code of \citep{hmss13}. The channel domain has periodic boundary conditions in the $x$ and $y$ directions, and assumes a rigid top and bottom.  
A third-order Runge-Kutta scheme with 
 adaptive time-stepping (CFL $= 0.9$) is used for time integration. Spatial discretization is based on pseudo-spectral decomposition using Fourier modes in the horizontal directions, and 2nd-order centered differences on a staggered grid in the vertical direction. 
 In our study, the domain size $L_x\times L_y\times H$ is
 $8000 \;\mbox{m} \times 8000 \;\mbox{m} \times 7000 \;\mbox{m}$ and the number of grid points is $256 \times 256 \times 400$,
 corresponding to horizontal and vertical grid spacings of $31.5$ m and $17.5$ m, respectively.
 
 The flow is confined to the channel by imposing zero vertical velocity $w$ on the top and bottom walls, respectively at $z=H$ and $z=0$.
The horizontal velocities $(u,v)$ are subject to the no-slip condition at the bottom and are allowed to slip at the top.  Altogether, $(u,v,w)$ satisfy the top and bottom boundary conditions:
\begin{equation}
    w|_{z=0,H} = 0,\quad
    u|_{z=0} = v|_{z=0} = 0,  \quad
    \frac{\partial u}{ \partial z} \bigg|_{z=H} = \frac{\partial v}{ \partial z} \bigg|_{z=H} = 0.
\end{equation}
The anomalous thermodynamic variables 
$f^\prime({\vec x},t)$ obey 
a homogeneous Neumann boundary conditions on both top and bottom boundaries, such that 
\begin{equation}
    \frac{\partial f^\prime}{ \partial z} \bigg|_{z=0,H}  = 0.
\end{equation}

The code from \cite{hmss13}
is adapted to our purpose of testing the inviscid conservation laws by using viscosity and diffusivity coefficients that are very small, with values $O(10^{-15}).$
As a comparison, results were compared for viscosity and diffusivity coefficients with values of zero and also values of $O(10^{-15})$ and $ O(10^{-9})$, and the differences  in $PV_u$ were small and of order of magnitude $O(10^{-15})$ and $ O(10^{-9})$, respectively, at measurement time $t=3$ min.  Given that the comparisons showed similar results, all results shown here are for the $O(10^{-15})$ case.

The background for the potential temperature $\theta$ is idealized to be a linear function of height: $\Tilde{\theta}(z) = \theta_0 + Bz,$ where $\theta_0 = 300 \; {\rm K}$ and $B = 3 \times 10^{-3} \; {\rm K \;m^{-1}}$. 
In this case, and using an ideal gas law together with the Clausius-Clapeyron relation, the saturation specific humidity $q_{vs}(z)$ may be approximated by 
\begin{equation}
q_{vs}(z) = \frac{q_{vs, 0}} {\Tilde{p}(z)/p_0} \exp \left( \frac{-L}{R_v} \left( \frac{1}{h(z) \tilde{\theta}(z) }- \frac{1}{\theta_0}\right) \right),
\label{eqn:clausius}
\end{equation}
where $q_{vs, 0} = 20 $ g kg$^{-1}$ is the value at $z= 0$, the latent heat $L \approx 2.5 \times 10^6$ J kg$^{-1}$ is constant, and $R_v \approx 462$ kg$^{-1}$ K$^{-1}$ is the gas constant for vapor \citep{hmss13}. The background pressure $\Tilde{p}(z)$ and the function $h(z)$ are given by
\begin{equation}
\frac{\tilde{p}(z)}{p_0} = \left( 1 - \frac{g}{Bc_p} \log \left( 1 + \frac{Bz}{\theta_0}\right) \right)^{c_p/R_d} = \left[ h(z) \right]^{c_p/R_d},
\label{eqn-qvs}
\end{equation}
where the (constant) specific heat is $c_p \approx 10^3$ J  kg$^{-1}$ K$^{-1}$ and the gas constant for dry air is $R_d \approx 287$ kg$^{-1}$ K$^{-1}$. The background specific humidity $\Tilde{q}_v(z)$ is chosen slightly below saturation by using formula \eqref{eqn:clausius} with $q_{vs, 0}$ replaced by $q_{v, 0} = 18$ g kg$^{-1}$.

A rising warm bubble is simulated in the domain to study $PV_u$ conservation in a moist atmosphere. The initial velocity is zero, while the initial anomalies of potential temperature and total water specific humidity are given by, respectively,
\begin{equation}
    \theta' (\Vec{x}, t=0) = A_{\theta} \cos^2\left(\frac{\pi r(\Vec{x})}{2r_c}\right)
\end{equation}
and
\begin{equation}
    q_t' (\Vec{x}, t=0) = A_{q} \cos \left(\frac{\pi(x-x_c)}{r_c}\right) \cos^2\left(\frac{\pi r(\Vec{x})}{2r_c}\right),
\end{equation}
where $r(\Vec{x}) = \sqrt{(x-x_c)^2 + (y-y_c)^2 + (z-z_c)^2}$, $x_c = L_x/2, \; y_c = L_y/2$, $z_c = 3H/7$, $r_c = 1.5$ km, $A_{\theta} = 2$ K and $A_q = 10^{-3}$ kg/kg.

\bibliographystyle{abbrv}

\bibliography{references}